\documentclass{ifacconf}

\usepackage{graphicx} 
\usepackage{natbib}
\usepackage{float}
\usepackage{caption}
\usepackage{subcaption}
\usepackage{textcomp}
\usepackage{xcolor}
\usepackage{amssymb,amsfonts}
\usepackage{algorithm} 
\usepackage{algpseudocode}
\usepackage{latexsym}
\usepackage{amsmath}
\usepackage{physics}
\usepackage{url}
\usepackage{soul}
\newcommand{\R}{\mathbb{R}}
\newcommand{\X}{\mathbf{X}}
\newcommand{\Y}{\mathbf{Y}}
\newcommand{\U}{\mathbf{U}}

\begin{document}
\begin{frontmatter}

\title{Data Driven Modeling of Turbocharger Turbine using Koopman Operator\thanksref{footnoteinfo}} 

\thanks[footnoteinfo]{This work was supported by Cummins Inc (GGM No. 20067719)}

\maketitle

\author[First]{Shrenik Zinage} 
\author[Second]{Suyash Jadhav}
\author[First]{Yifei Zhou}
\author[First]{Ilias Bilionis}
\author[First]{Peter Meckl}

\address[First]{School of Mechanical Engineering, Purdue University, West Lafayette, IN-USA (e-mail: szinage@purdue.edu, zhou849@purdue.edu, ibilion@purdue.edu, meckl@purdue.edu)}

\address[Second]{Battery Controls Engineer at Canoo, Torrance, CA-USA (e-mail: suyash.jadhav@canoo.com)}

\begin{abstract}                
A turbocharger plays an essential part in reducing emissions and increasing the fuel efficiency of road vehicles. The pulsating flow of exhaust gases, along with high heat exchange from the turbocharger casing, makes developing control-oriented models difficult. Several researchers have used maps provided by manufacturers to solve this problem. These maps often fail to incorporate any heat transfer effects and are unsuitable for wide operating regions. Also, with the availability of more and better sensor data, there is a need for a method that can exploit this to obtain a better predictive model. Koopman approaches rely on the observation that one can lift the nonlinear dynamics of the turbine into an infinite-dimensional function space over which dynamics are linear. The objective of this paper is to develop a model to predict the transient and steady-state behavior of the turbine using the Koopman operator which can be helpful for control design and analysis. Our approach is as follows. We use experimental data from a Cummins heavy-duty diesel engine to develop a turbine model using Extended Dynamic Mode Decomposition, which approximates the action of the Koopman operator on a finite-dimensional subspace of the space of observables. The results demonstrate superior performance compared to a tuned nonlinear autoregressive network with an exogenous input model widely used in the literature. The performance of these two models is analyzed based on their ability to predict turbine transient and steady-state behavior.

\end{abstract}


\end{frontmatter}

\section{Introduction}

\begin{figure}[htbp]
	\centering
      \includegraphics[width=8.4cm]{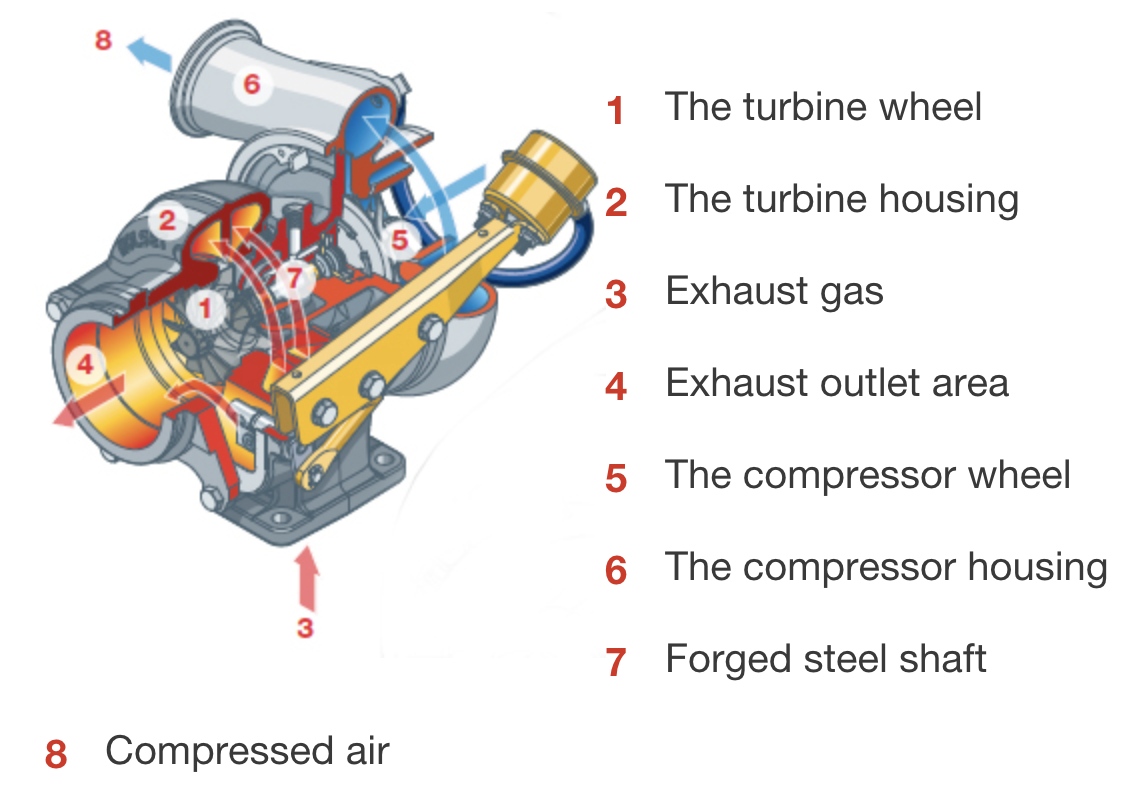}
      \caption{Turbocharger Schematic \citep{cummins}}
	\label{fig:turbocharger}
\end{figure}

A turbocharger plays a vital role in reduction of emissions and fuel consumption of road vehicles. It consists of an exhaust gas turbine and a compressor coupled together by a shaft as shown in Fig. \ref{fig:turbocharger}. The turbine extracts energy from the gas by rotating the turbine wheel which in turn drives the compressor. The pulsating flow of exhaust gases with the high heat transfers from the turbocharger casing makes the dynamical modeling of turbochargers very challenging.

The methods for modeling the turbine subsystem have improved over time. Detailed models incorporating energy transformation and losses within every element from a variable geometry turbine can be found in \cite{decombes2002simulation}. A good background on flow and heat transfer in a turbocharger under both gas stand (continuous flow) and engine-like (pulsating flow) conditions is presented in \cite{lim2016flow}. \cite{kachele2020turbocharger} discusses a computational fluid dynamics (CFD) approach towards simulating the turbine in engine simulations. \cite{galindo2014development} provides a 1-dimensional variable geometry turbine model incorporating pulsating flows. The most common approach to model turbocharger dynamics is by using maps provided by manufacturers. Several researchers have proposed modeling techniques for turbine and compressor using polytropic, isentropic, and monotonic maps \citep{serrano2008model,eriksson2002modeling,chauvin2006experimental}. Often, these maps cannot fully capture the nonlinear dynamics, especially during transient load applications. A 0-dimensional model that predicts and analyzes the engine's thermodynamic properties along with the turbine maps for model-based control is provided by \cite{moraal1999turbocharger}. As these maps are developed on a test bench with ideal boundary and initial conditions, they fail to incorporate any heat transfer effects \citep{marelli2017heat} and are not suitable for wide operating regions. Due to the higher computational requirement and the complexity of all these physics-based models, mean value models for capturing gas exchange dynamics have also been used \citep{jung2005calibratable,wang2008hybrid}. 

Various techniques have also been used in the literature to extract information from data while maintaining robustness by the use of physics-based models. For instance, \cite{kumar2019hybrid} used a hybrid modeling technique to simulate the behaviour of a diesel engine by switching between an empirical model and a physics-based model. A hybrid approach by interpolating between white and black box models can be found in \cite{moraal1999turbocharger}. In \cite{kushwaha2015air}, a nonlinear autoregressive network with exogenous input (NARX) is used to identify the air path of a turbocharged diesel engine. A data-driven model of the turbocharger is developed in \cite{pulpeiro2019modeling} to predict pressure and temperature ratios over the compressor. \cite{asgari2014modeling} use a NARX architecture to model the turbine speed and turbocharger outlet temperatures. Neural networks are also used to improve the performance of dynamic models by modeling heat transfer from the turbine as shown by \cite{huang2018applying} where the difference between map-based prediction and measurement data of turbine outlet temperature is modeled using a neural network. 

The Koopman operator is a linear infinite-dimensional operator that converts a nonlinear finite-dimensional system into a linear system in a Koopman invariant subspace. In other words, it is a linear infinite-dimensional operator that describes the evolution of observables which are functions of states. The conventional linearization techniques are only able to approximate the nonlinear dynamics over a local region. However, the Koopman operator allows one to fully capture the governing nonlinear dynamics over a much broader state space but comes with the downside that it is computationally infeasible. Towards this aim, recent methods have used extended dynamic mode decomposition (EDMD) \citep{williams2015data_koopman_accuracy_1} or neural networks to obtain a finite approximation of this operator.
In addition, these methods allow us to use tools from linear systems theory to analyse a nonlinear system. For instance, it can be used to design a linear controller, which can then be applied to the original nonlinear system of interest ~\citep{brunton2016koopman_control_1,proctor2018generalizing_koopman_control_3,williams2016extending_koopman_control_4}.
An extension of these methods for controlled systems can be found in \cite{proctor2016dynamic, korda2018linear, kaiser2020data, ma2019optimal}.
The major challenge in the EDMD approach lies in selecting the correct set of observables. Various studies have looked at using different dictionary of functions such as radial basis functions (RBFs)\citep{korda2018linear} and polynomial basis functions\citep{cibulka2019data} as observables to approximate the nonlinear dynamics. However, a systematic method for selecting these observables that best mimic the Koopman operator is still a work in progress.
Some recent studies to address the above issue using neural networks, trial and error methods, and analytical methods can be found in \citep{lusch2018deep_machine_learning,yeung2019learning,abraham2019active_trial,mamakoukas2021derivative}.

The high degree of nonlinearity in the turbine, particularly due to heat transfer, combined with a strong coupling between the exhaust gas recirculation (EGR) and variable geometry turbocharger (VGT) systems, makes developing control oriented models difficult. Hence, this paper aims to develop a control oriented model to predict the turbine's transient and steady-state behavior using the Koopman operator.

The main contributions of this paper are as follows. First, this paper introduces a data-driven approach to model the turbine outputs, namely: the turbine speed ($N_t$), and turbine outlet temperature ($T_\text{tur,out}$) using the Koopman operator. Second, the accuracy of this model in predicting the transient and steady state behaviour is compared with a tuned NARX architecture. 

We have organized our paper as follows. Section~\ref{sec:tt} discusses the mathematical formulation of the turbine. Koopman operator is discussed in Section~\ref{sec:koopman}. Section~\ref{sec:narx} describes the NARX approach used for comparison. The results are provided in Section~\ref{sec:results}, and the conclusions are presented in Section~\ref{sec:conclusions}.

As this paper is a part of collaborative work sponsored by Cummins Inc, the data is collected on a Cummins heavy duty diesel engine and the plots shown in this paper are normalized in accordance with Cummins guidelines.

\section{Turbocharger Turbine}
\label{sec:tt}

A turbocharger is used to extract the energy obtained from exhaust gases into useful work by rotating the turbine wheel. Since it is an open system it can exchange both matter and energy with its surroundings. The energy equation is given by:
\begin{equation}
\frac{dE}{dt} = \dot{H}_\text{in} - \dot{H}_\text{out} - \dot{W}_\text{turbine} + \dot{Q}_\text{tur,housing},
\end{equation}
where $\dot{H}_\text{in}$ and $\dot{H}_\text{out}$ denote the rate of change in enthalpy at turbine inlet and outlet with respect to time, $\dot{W}_{\text{turbine}}$ denotes the rate of change in the work done by the turbine with respect to time, and $\dot{Q}_\text{tur,housing}$ indicates the rate of heat transfer from the turbine. Since the turbine is neither a heat sink nor a heat source, we can assume $\frac{dE}{dt} = 0$. Therefore, we have:
\begin{equation}
    \dot{H}_\text{in} =  \dot{H}_\text{out} + \dot{W}_\text{turbine} - \dot{Q}_\text{tur,housing}.
\end{equation}
If $\dot{W}_t$ is the mass flow rate of exhaust gas through the turbine, $c_p$ is isobaric specific heat capacity, and $\tau$ is the torque generated by the turbine, it can be simplified as:
\begin{equation}
    \dot{W}_tc_pT_\text{tur,in} = \dot{W}_tc_pT_\text{tur,out} + \tau N_t - \dot{Q}_\text{tur,housing},
\end{equation}
where $T_\text{tur,in}$, and $T_\text{tur,out}$ denote the turbine inlet and outlet temperatures, respectively. The turbocharger speed $N_t$ can modeled using Newton's second law for rotating systems with an additional friction term as shown below:
\begin{equation}
    \frac{dN_t}{dt} = \frac{1}{J_t}\left[\frac{P_t}{N_t} - \frac{P_c}{N_t} - M_\text{fric}(N_t)\right],
    \label{eq:n_t}
\end{equation}
where $J_t$ is the rotational inertia of the turbine, $M_\text{fric}$ is the friction component which is a function of $N_t$, and $P_t$, $P_c$ denote the turbine and compressor powers, respectively. 

In order to model two significant outputs from the turbine namely, $N_t$ and $T_\text{tur,out}$, the following inputs are considered: 
\begin{itemize}
    \item $u_\text{vgt}$ - Actuation of VGT (\%)
    \item $u_\text{egrv}$ - Valve actuation of EGR (\%)
    \item $T_\text{tur,in}$ - Turbine inlet temperature (K)
    \item $P_\text{tur,in}$ - Turbine inlet pressure (kPa)
    \item $N_\text{e}$ - Engine speed (rpm)
    \item $m_\text{f}$ - Mass of the fuel injected (mg)
    \item $T_\text{oil}$ - Oil temperature (K)
    \item $T_\text{coolant}$ - Coolant temperature (K)
    \item $T_\text{comp,out}$ - Compressor outlet temperature (K)
\end{itemize}

These inputs are selected based on the physical intuition of the dynamic behavior of the system. As the thermal dynamics for modeling $T_\text{tur,out}$ are caused by fuel burned in the combustion chamber, $m_\text{f}$ is included as an input alongside $T_\text{tur,in}$. We have also added $T_\text{oil}$ as an input to incorporate the influence of heat transfer from the turbine as the external disturbance $Q_\text{tur,housing}$ is the major cause of nonlinearity in the system which is challenging to measure. It is also found that by adding $T_\text{coolant}$, and $T_\text{comp,out}$, there is a slight improvement in the prediction performance. Hence, they are also considered as inputs in this paper.

\section{The Koopman Operator}
\label{sec:koopman}
\subsection{Introduction to Koopman Operator}

Consider a continuous-time nonlinear system given by
\begin{align}
    \dot{x}=f(x,u),\quad x(t_0)=x_0,
    \label{eqn:nonlinear_system}
\end{align}
where $x$ in $\R^n$ is the state, and $u\in\mathbb{R}^m$ is the input. Let $F$ be the flow that maps $x(t)$ to the solution at time $t+\Delta t$ as shown below:
\begin{align}
   {F}\left({x}\left(t\right),{u}\left(t\right),\Delta t\right)={x}\left(t_{0}\right)+\int_{t}^{t+\Delta t} {f}({x}(\tau),u(\tau)) d \tau,
\end{align}
where $f$ is Lipschitz continuous. In a discrete sense with time step $\Delta t_d$, it can be stated as $x_{l+1}=F((x_{l},u_{l}),\Delta t_d)$ where $x_l:=x(t_l)$, and $t_l\in[l\Delta t_d,(l+1)\Delta t_d]$. Let the function that evolves this dynamics of discretized system be given by: 
\begin{align}
    x_{l+1} = g(x_{l},u_{l}).
    \label{eq:discrete}
\end{align}
The Koopman operator $\mathcal{K}$ is a linear infinite dimensional operator that advances the measurement functions with the flow $F$ as follows:
\begin{align}
    \mathcal{K}\boldsymbol{\psi}=\boldsymbol{\psi}\circ F,
\end{align}
where $\boldsymbol{\psi}$ is an element in the infinite dimensional Hilbert space (also known as observable functions). Since the Koopman operator is infinite dimensional, it is computationally infeasible for control applications. Hence, finite-dimensional approximations of this operator via methods like EDMD, and neural networks have received increased attention in the controls community and is illustrated in the following section.

\subsection{Finite approximations of the Koopman operator}
By exact discretization, the Koopman operator $\mathcal{K}_d$ for a nonlinear discrete system associated with system (\ref{eq:discrete}) can be stated in terms of $\mathcal{K}$ and the sampling time $\Delta t_d$ as follows:
\begin{align}
     \mathcal{K} = \frac{\text{log}(\mathcal{K}_d)}{\Delta t_d}.
\end{align}
Let $\mathcal{O}$ be the space of complex valued observable functions $\psi:\mathbb{R}^n\rightarrow\mathbb{R}$. The Koopman operator $\mathcal{K}_d: \mathcal{O} \rightarrow \mathcal{O}$ is defined by:
\begin{equation}
\mathcal{K}_d \boldsymbol{\psi}\left(x_{l},u_{l}\right)=\boldsymbol{\psi}\left(g(x_{l},u_{l})),u_{l+1}\right)=\boldsymbol{\psi}\left(x_{l+1}, u_{l+1}\right).
\end{equation}
The observable function $\boldsymbol{\psi}(x)$ can be written as:
\begin{equation}
\boldsymbol{\psi}(x):=\left[\psi_1(x),\;\psi_2(x),\dots {\psi}_{N_l}(x)\right]^\mathrm{T},
\end{equation}
where $\psi_i(x):\mathbb{R}^n\rightarrow\mathbb{R},\;\forall\;i\in\{1,\dots, N_l\}$, $N_l\gg n$. The state $z$ represents the system's states in lifted space and is given by:
\begin{equation}
z_l=\boldsymbol{\psi}(x_l),
\end{equation}
where $\boldsymbol{\psi}(\cdot)$ is defined as the lifting mapping.

\subsection{Extended Dynamics for Turbocharger Turbine}

This section discusses the principal steps for approximating the nonlinear turbine dynamics to a higher dimensional linear state space model based on the Koopman operator. This linear model can be represented as follows:
\begin{equation}
\begin{split}
z_{l+1} &= Az_{l}+Bu_{l}\\
\hat{y}_l &= Cz_{l},\\
\end{split}
\end{equation}
where $\hat{y}_l$ is the output estimate, $z$ is referred to as the lifted state, $N_l$ is the dimension of $z$, $A \in \mathbb{R}^{N_l \times N_l}$, $B \in \mathbb{R}^{N_l \times m}$ ($m$ is the dimension of the inputs), $C \in \mathbb{R}^{k \times N_l}$ ($k$ is the dimension of output estimate), $u_l\in\mathbb{R}^m$, and $l=0,1,\dots,N-1$. Here, $N$ indicates the total number of data points. The initial lifted state $z_0$ is given by:
\begin{equation}
z_0=\boldsymbol{\psi}(x_0):=\left[\psi_1(x_0)),\;\;\;\psi_2(x_0)),\dots, {\psi}_{N_l}(x_0)\right]^\mathrm{T}.
  \label{eqn:koopman_initial_condition}
\end{equation}

\subsection{Extended Dynamic Mode Decomposition approach to compute $A$ and $B$}

We will discuss an empirical method to calculate the matrices  $A$ and $B$ in this section. Let the input be $u_l$ that will take the states of the turbine (i.e, $N_t$ and $T_\text{tur,out}$) from $x_l$ to $x_{l+1}$. By doing this we will be able to form the data matrices $\X$, $\Y$, and $\U$ as shown below:
\begin{subequations}
\begin{align}
    &\X := [x_0,.....x_{N-1}]\\
    &\Y := [x_1,.....x_N]\\
    &\U := [u_0,.....u_{N-1}].
\end{align}
\label{eq:matrices}
\end{subequations}
Now, provided we have $\X$, $\Y$ and $\U$, the matrices $A$ and $B$ are computed using the solution to the minimization problem:
\begin{equation}
\min _{A, B}\left\|\Y_{\mathrm{lift}}-A \X_{\mathrm{lift}}-B\U\right\|,
 \label{eqn:least_squares_optimization}
\end{equation}
where $\|\cdot\|$ is the square root of the sum of absolute squares of its elements, and  $\X_\mathrm{lift}$ and $\Y_\mathrm{lift}$ are given by:
\begin{equation}
	 \begin{split}
		&\X_{\mathrm{lift}}:=\left[\boldsymbol{\psi}(\boldsymbol{x}_0), \ldots, \boldsymbol{\psi}(\boldsymbol{x}_{N-1})\right]\\ 
		&\Y_{\mathrm{lift}}:=\left[\boldsymbol{\psi}(\boldsymbol{x}_{1}), \ldots, \boldsymbol{\psi}(\boldsymbol{x}_N)\right].\\
	\end{split}
\end{equation}
The analytical solution to \eqref{eqn:least_squares_optimization} is as shown below:
\begin{equation}
[A, B]=\boldsymbol{\Y}_{\text {lift}}\left[\begin{array}{c}
\X_{\mathrm{lift}} \\
\boldsymbol{\U}
\end{array}\right]^{\dagger},
\label{eqn:regression}
\end{equation}
where $\dagger$ denotes the pseudoinverse of a matrix.

\section{NARX Approach}
\label{sec:narx}
The NARX is a recurrent dynamic network that uses a feedforward network and feedback information in the form of time-delayed inputs \emph{$u_{l-1}$}, or output estimates \emph{$y_{l-1}$} to predict the output $y_l$. This prediction $y_l$ is given by:
\begin{equation}
    y_l = h(y_{l-1}, ..,y_{l-n_y},u_{l-1}, ..,u_{l-n_u}),
\end{equation}
where $u$ are inputs to the network, and $n_u$, $n_y$ are timestep delays at input and output predictions, respectively. 

One of the major advantages of using this network is that it acts as an adaptive filter which eliminates the high frequency components from the predictions \citep{mandic2001recurrent}. This is an added advantage for modeling $T_\text{tur,out}$ where it is certain that the thermal dynamics cannot be found at higher frequencies, and hence can be filtered \citep{suyash2021}.
In order to make a fair comparison with the EDMD approach, a first order Markov property is assumed as shown below:
\begin{equation}
    y_l = h_\text{NARX}(y_{l-1},u_{l-t_d}).
\end{equation}
where $t_d$ indicates the time delay of inputs.
Due to physical location and operating environment, a thermocouple with large thermal mass is used to measure $T_\text{tur,out}$. This results in a significant delay in the temperature measurement. Thus, different input delays are assumed for modeling respective outputs (i.e, $N_t$ and $T_\text{tur,out}$).

\begin{figure}[htbp]
	\centering
      \includegraphics[width=8.4cm]{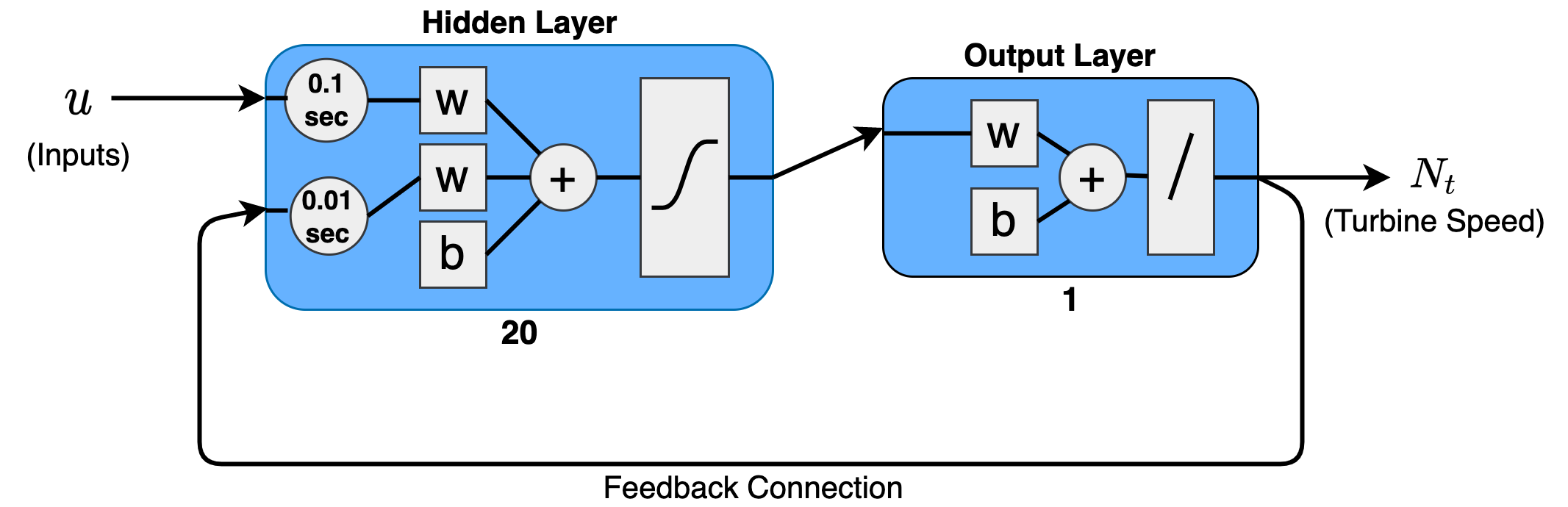}
      \caption{Tuned NARX architecture for modeling $N_t$}
	\label{fig:NARX_Nt}
\end{figure}

Since there was an instantaneous response in $N_t$ when the inputs are excited, the input delay is set to $0.1$ s \citep{suyash2021}. The number of neurons in the hidden layer is explored from 4 to 20 with an increment of 2, and the best performance is obtained when the number of neurons is set to 20. Fig. \ref{fig:NARX_Nt} shows the tuned NARX architecture for modeling $N_t$.

\begin{figure}[htbp]
	\centering
      \includegraphics[width=8.4cm]{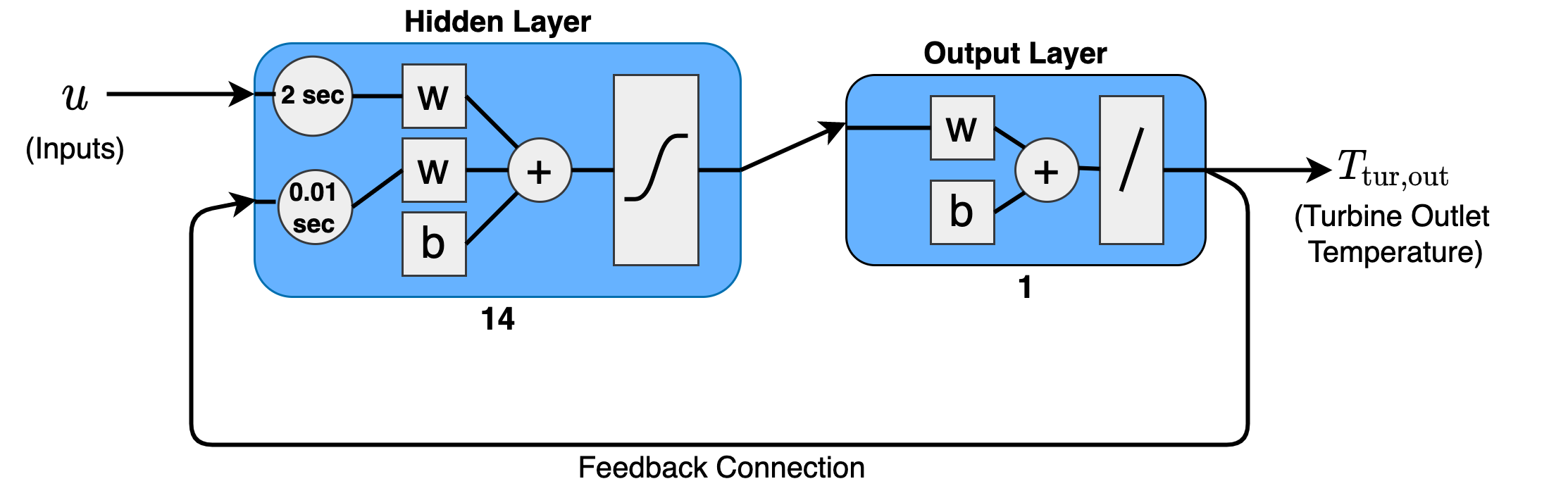}
      \caption{Tuned NARX architecture for modeling $T_\text{tur,out}$}
	\label{fig:NARX_T_tur_out}
\end{figure}

Due to large delay in thermocouple measurement of the temperature, an input delay of 2 s is used to model $T_\text{tur,out}$ \citep{suyash2021}. The number of neurons is again explored from 4 to 20 with an increment of 2, and the best performance is obtained for 14 neurons. Fig. \ref{fig:NARX_T_tur_out} shows the tuned NARX architecture for modeling $T_\text{tur,out}$.

For both models, the feedback delay is assumed to be 0.01 s. The data sampling frequency is set to 100 Hz, and the network is trained using a Levenberg-Marquardt backpropagation algorithm (also called damped least-squares) with Bayesian regularization. The tuned hyperparameters for modeling these two outputs is summarized in Table \ref{tab:narx_hyp}.
\begin{table}[htbp]
\begin{center}
\captionsetup{width=8.5cm}
\caption{Tuned hyperparameters for NARX architecture}
\label{tab:narx_hyp}
\begin{tabular}{cccc}
      \hline
	\textbf{Output} & \textbf{Input Delay} & \textbf{Feedback Delay} & \textbf{Neurons}\\
      \hline
      $N_t$ & 0.1 s & 0.01 s & 20\\
      $T_\text{tur,out}$ & 2 s & 0.01 s & 14\\
	\hline
    \end{tabular}
\end{center}
\end{table}

\section{Results}
\label{sec:results}

\subsection{Data Generation}                
Transient experimental data from heavy duty diesel engine was provided by Cummins Inc. Multiple test cycles were created in which the inputs were manipulated using different excitation signals. This dataset covered the entire operating region of this engine and the sampling frequency was 100 Hz. The performance of the trained models using both NARX and EDMD approaches is then tested on two different validation datasets namely: FTP and SET duty cycles to analyze the efficacy of these models in predicting the transient and steady state behavior of the turbine. In this paper, the experimental data is considered to be the ground truth and prediction results are compared with it.

\subsection{Metrics}

The performance metrics used are Normalized Root Mean Square Error (NRMSE), R-squared ($\text{R}^2$), and Mean Absolute Percent Error (MAPE) as shown below:
\begin{equation}
   \text{NRMSE}=\frac{\sqrt{\frac{1}{N}\sum_{i=1}^{N}e_i^2}}{mean(\hat y)} 
   \label{myeq3}
\end{equation}
\begin{equation}
\text{R}^{2}=1-\frac{\sum_{i=1}^{N}e_i^2}{\sum_{i=1}^{N}(y_i - mean(y_i))^2}
\label{myeq4} 
\end{equation}
\begin{equation}
   \text{MAPE}=\frac{1}{N}\sum_{i=1}^{N}\frac{|e_{i}|}{|y_{i}|}.100
   \label{myeq5},
\end{equation}
where $e_{i}$ is the error between predicted value ($\hat y_{i}$) and measured value ($y_{i}$), and $N$ is the total number of data points. Throughout this paper, these performance metrics are used to compare the predictions with measured signals to evaluate the model accuracy.

\subsection{Data collection, Lifting and Regression using EDMD}
Firstly, the matrices $\X$, $\Y$, and $\U$ as per \eqref{eq:matrices} are computed using the actual inputs and normalized states ($\mu = 0,\sigma = 1$) obtained from the experimental data. Once the data are collected, the lifted state $z$ is calculated by concatenating the actual states of the turbine (i.e. $N_\text{t}$ and $T_\text{tur,out}$) with ``polyharmonic radial basis functions'' with different centers as shown below:
\begin{equation}
z = [x,  \sqrt{\alpha_i}\log(\sqrt{\alpha_i})]^T, \hspace{5mm} \alpha_i=\sum_{j=1}^2(\boldsymbol{x}(j)-\boldsymbol{c}_i(j))^2,
\end{equation}
where $c_i$ are the radial basis function (RBF) centers which is a random vector taken from a uniform distribution over $[-1.8,1.8]^2$ (as the normalized states of $N_t$ and $T_\text{tur,out}$ were varying between -1.8 and 1.8), for $i  \in [0, 150]_d$. Unfortunately, there is no universal guideline for determining which basis functions to use for a given system. It is all about trial and error where we try using different basis functions like polynomial basis functions and RBF's (gaussian, multiquadric, inverse quadric, inverse multiquadric, and polyharmonic spline). For our case, the polyharmonic RBF yielded us with the best results. Once the lifting is done, the regression is performed as per \eqref{eqn:regression} to get the system matrix $A$ and input matrix $B$. These matrices $A$ and $B$ are then used to predict the turbocharger turbine responses for FTP and SET duty cycles. In order to obtain the actual data values of the outputs, the prediction at each sampling time is denormalized with the help of the mean and standard deviation of the training data.

\begin{table}[htbp]
\begin{center}
\captionsetup{width=8.5cm}
\caption{Prediction performance of $N_\text{t}$ using EDMD}
\label{tab:n_t_results}
\begin{tabular}{cccc}
      \hline
	\textbf{No. of RBF} & \textbf{NRMSE} & \textbf{$\text{R}^2$} & \textbf{MAPE}\\
	\hline
	\multicolumn{4}{c}{\textbf{FTP Duty Cycle}} \\
      \hline
       0 (no lift) & 0.147 & 0.946 & 16.92\\
      50 & 0.111 & 0.970 & 8.96\\
      \textbf{100} & 0.110 & 0.971 & 8.92\\
      150 & 0.111 & 0.970 & 8.77\\
	  \hline
	\multicolumn{4}{c}{\textbf{SET Duty Cycle}} \\
	\hline
       0 (no lift) & 0.08 & 0.958 & 9.82\\
      50 & 0.02 & 0.997 & 2.70 \\
      \textbf{100} & 0.02 & 0.998 & 2.44\\
      150 & 0.03 & 0.995 & 4.18\\
	\hline
    \end{tabular}
\end{center}
\end{table}

\begin{table}[htbp]
\begin{center}
\captionsetup{width=8.5cm}
\caption{Prediction performance of $T_\text{tur,out}$ using EDMD}
\label{tab:t_tur_out_results}
\begin{tabular}{cccc}
      \hline
	\textbf{No. of RBF} & \textbf{NRMSE} & \textbf{$\text{R}^2$} & \textbf{MAPE}\\
	\hline
	\multicolumn{4}{c}{\textbf{FTP Duty Cycle}} \\
      \hline
       0 (no lift) & 0.053 & 0.55 & 4.57\\
      50 & 0.030 & 0.85 & 2.63\\
      \textbf{100} & 0.029 &  0.90 & 2.15 \\
      150 & 0.034 & 0.82 & 2.93 \\
      \hline
	\multicolumn{4}{c}{\textbf{SET Duty Cycle}} \\
	\hline
       0 (no lift) & 0.060 & 0.55 & 5.08 \\
      50 & 0.045 & 0.75 & 3.67\\
      \textbf{100} & 0.044 & 0.76 &  3.45 \\
      150 & 0.069 & 0.40 & 4.83\\
	\hline
    \end{tabular}
\end{center}
\end{table}

From Table \ref{tab:n_t_results} and Table \ref{tab:t_tur_out_results}, it can be seen that as the number of basis functions increases, the modeling accuracy for $N_\text{t}$ and $T_\text{tur,out}$ increases and then starts to decrease after a particular number of basis functions, which is 100. This might be caused due to the overfitting of the training data. Since the optimal prediction performance is achieved when the number of radial basis functions is set to 100, the dimension of the extended states will be 102 (i.e, $N_\text{t}$ and $T_\text{tur,out}$ $+$ number of RBF's).

\subsection{Comparison with NARX model}

\begin{figure}[htbp]
	\centering
      \includegraphics[width=8.2cm]{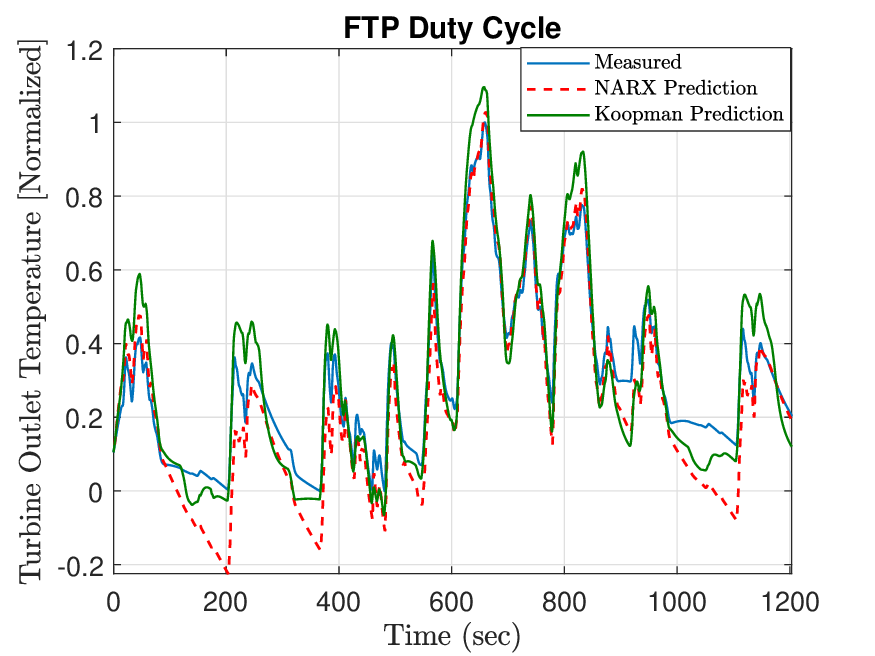}
      \caption{Turbine outlet temperature for FTP dataset}
	\label{fig:T_tur_out_hftp}
\end{figure}

\begin{figure}[htbp]
	\centering
      \includegraphics[width=8.2cm]{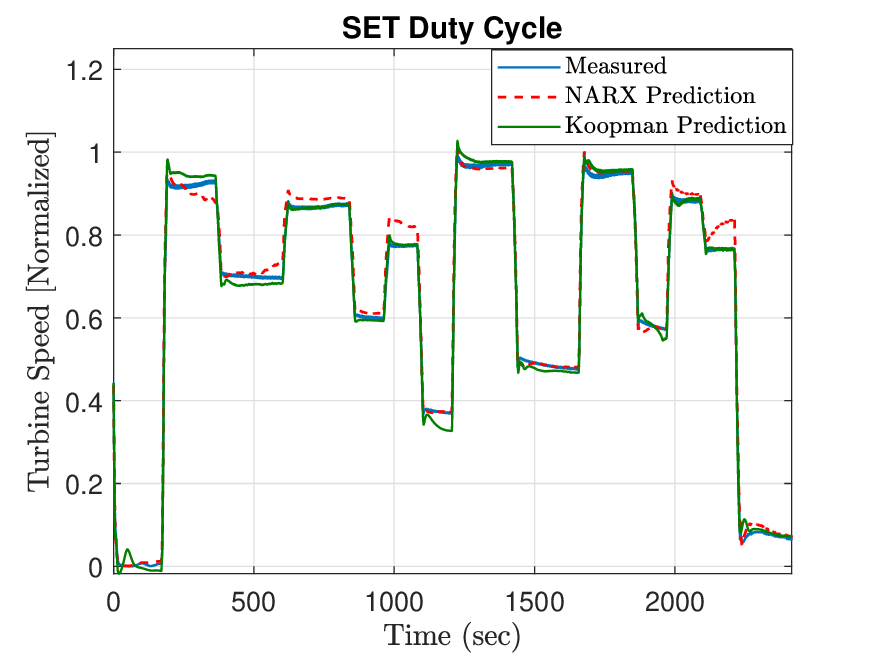}
      \caption{Turbine speed for SET dataset}
	\label{fig:N_t_rmcset}
\end{figure}

\begin{figure}[htbp]
	\centering
      \includegraphics[width=8.2cm]{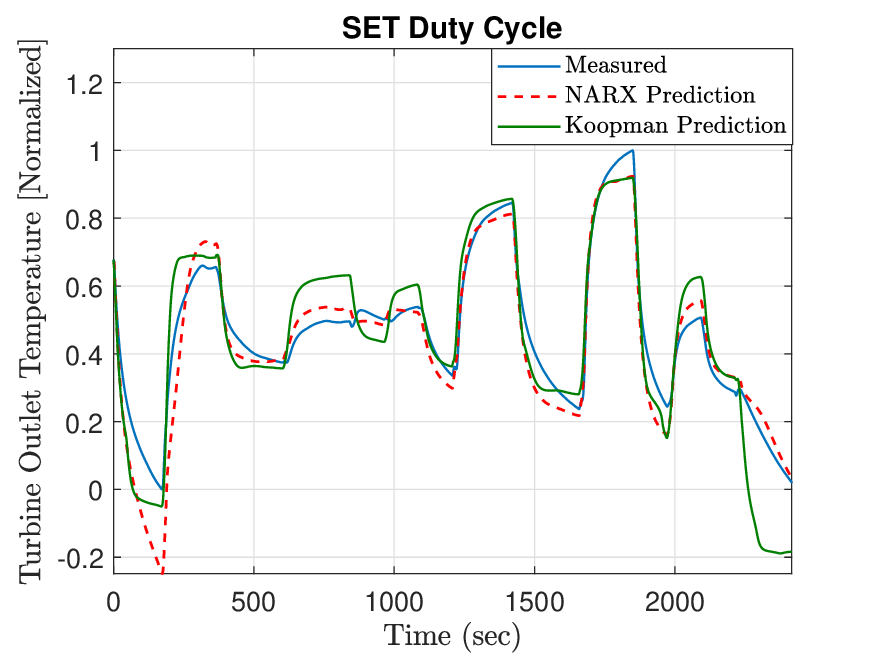}
      \caption{Turbine outlet temperature for SET dataset}
	\label{fig:T_tur_out_rmcset}
\end{figure}

Table \ref{tab:hftp_results} and Table \ref{tab:rmcset_results} show the model accuracy of the turbine outputs of both methods in FTP and SET test cycles. It can be seen that the EDMD approach is able to provide a better model accuracy in predicting both $N_t$ and $T_\text{tur,out}$ in the FTP cycle and $N_t$ in the SET cycle. However, the NARX model seems to provide a better MAPE value than the Koopman model while modeling $N_t$ in both the duty cycles and $T_\text{tur,out}$ in the SET cycle. This is indicative that the NARX model is able to handle outliers in the data better than the Koopman model in most of the scenarios. Fig. \ref{fig:T_tur_out_hftp} shows the comparison plot for predicting $T_\text{tur,out}$ in FTP duty cycle. It can be observed that the Koopman model is able to predict $T_\text{tur,out}$ at low turbine speeds with greater accuracy. However, the NARX model is able to provide a better prediction at higher turbine speeds. From Fig. \ref{fig:N_t_rmcset}, it can be seen that the Koopman model is able to outperform the NARX model in modeling $N_t$ in the SET cycle. However, when it comes to modeling $T_\text{tur,out}$, the NARX model shows a better prediction as compared to the Koopman model (Fig. \ref{fig:T_tur_out_rmcset}).



\begin{table}[htbp]
\begin{center}
\captionsetup{width=8.5cm}
\caption{Model accuracy for FTP Dataset}
\label{tab:hftp_results}
\begin{tabular}{cccc}
      \hline
	\textbf{Method} & \textbf{NRMSE} & \textbf{$\text{R}^2$} & \textbf{MAPE}\\
	\hline
	\multicolumn{4}{c}{\textbf{Turbine Speed ($N_\text{t}$)}} \\
      \hline
       NARX & 0.12 & 0.96 & 7.20\\
      EDMD ($N_\text{RBF}$ = 100) & 0.11 & 0.97 & 8.92\\
      \hline
	\multicolumn{4}{c}{\textbf{Turbine Outlet Temperature ($T_\text{tur,out}$})} \\
	\hline
       NARX & 0.03 & 0.81 & 2.91\\
      EDMD ($N_\text{RBF}$ = 100) & 0.029 & 0.9 & 2.15\\
	\hline
    \end{tabular}
\end{center}
\end{table}

\begin{table}[htbp]
\begin{center}
\captionsetup{width=8.5cm}
\caption{Model accuracy for SET Dataset}
\label{tab:rmcset_results}
\begin{tabular}{cccc}
      \hline
	\textbf{Method} & \textbf{NRMSE} & \textbf{$\text{R}^2$} & \textbf{MAPE}\\
\hline
	\multicolumn{4}{c}{\textbf{Turbine Speed ($N_t$)}} \\
      \hline
       NARX & 0.03 & 0.99 & 2.30\\
      EDMD ($N_{RBF}$ = 100) & 0.02 & 0.99 & 2.44\\
      \hline
	\multicolumn{4}{c}{\textbf{Turbine Outlet Temperature ($T_\text{tur,out}$) }} \\
	\hline
       NARX & 0.03 & 0.89 & 2.29\\
      EDMD ($N_{RBF}$ = 100) & 0.044 & 0.76 & 3.45\\
	\hline
    \end{tabular}
\end{center}
\end{table}

\begin{figure}[htbp]
	\centering
      \includegraphics[width=8.2cm]{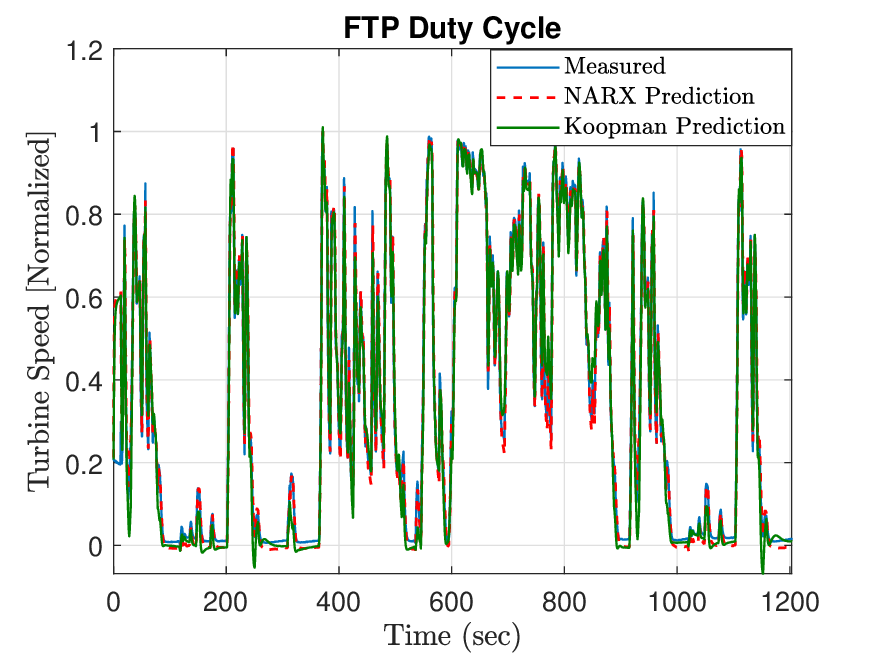}
      \caption{Turbine speed for FTP dataset}
	\label{fig:N_t_hftp}
\end{figure}

\section{Conclusion}
\label{sec:conclusions}

This paper presents a data-driven control-oriented model of a turbocharger turbine using the Koopman Operator. This model is linear and can be used for control design and analysis. It is compared against a tuned NARX network and tested on two different datasets, namely: FTP and SET duty cycles to validate the efficacy of both these methods in predicting the transient and steady state behavior of the turbine. The results demonstrate superior performance of the EDMD approach in modeling the turbine speed for both test cycles. When it comes to modeling the turbine outlet temperature ($T_\text{tur,out}$) in the FTP duty cycle, the Koopman model is able to predict $T_\text{tur,out}$ with a better accuracy whereas the NARX model showed superior accuracy in the SET duty cycle. 

\section{Acknowledgement}

We would like to sincerely thank Akash Desai from Cummins Inc for his valuable feedback and guidance at all stages of the project. The authors would also like to thank Dr. Lisa Farrell and Clay Arnett from Cummins Inc for sponsoring this research and providing us with the technical input and experimental data to run the simulations.

\bibliography{main.bib}         

\begin{thebibliography}{32}
\providecommand{\natexlab}[1]{#1}
\providecommand{\url}[1]{\texttt{#1}}
\providecommand{\urlprefix}{URL }
\expandafter\ifx\csname urlstyle\endcsname\relax
  \providecommand{\doi}[1]{doi:\discretionary{}{}{}#1}\else
  \providecommand{\doi}{doi:\discretionary{}{}{}\begingroup
  \urlstyle{rm}\Url}\fi

\bibitem[{Abraham and Murphey(2019)}]{abraham2019active_trial}
Abraham, I. and Murphey, T.D. (2019).
\newblock Active learning of dynamics for data-driven control using koopman
  operators.
\newblock \emph{IEEE Transactions on Robotics}, 35(5), 1071--1083.

\bibitem[{Asgari et~al.(2014)Asgari, Venturini, Chen, and
  Sainudiin}]{asgari2014modeling}
Asgari, H., Venturini, M., Chen, X., and Sainudiin, R. (2014).
\newblock Modeling and simulation of the transient behavior of an industrial
  power plant gas turbine.
\newblock \emph{Journal of Engineering for Gas Turbines and Power}, 136(6).

\bibitem[{Brunton et~al.(2016)Brunton, Brunton, Proctor, and
  Kutz}]{brunton2016koopman_control_1}
Brunton, S.L., Brunton, B.W., Proctor, J.L., and Kutz, J.N. (2016).
\newblock Koopman invariant subspaces and finite linear representations of
  nonlinear dynamical systems for control.
\newblock \emph{PloS one}, 11(2), e0150171.

\bibitem[{Chauvin et~al.(2006)Chauvin, Corde, Petit, and
  Rouchon}]{chauvin2006experimental}
Chauvin, J., Corde, G., Petit, N., and Rouchon, P. (2006).
\newblock Experimental motion planning in airpath control for hcci engine.
\newblock In \emph{2006 American Control Conference}, 6--pp. IEEE.

\bibitem[{Cibulka et~al.(2019)Cibulka, Hani{\v{s}}, and
  Hrom{\v{c}}{\'\i}k}]{cibulka2019data}
Cibulka, V., Hani{\v{s}}, T., and Hrom{\v{c}}{\'\i}k, M. (2019).
\newblock Data-driven identification of vehicle dynamics using koopman
  operator.
\newblock In \emph{2019 22nd International Conference on Process Control
  (PC19)}, 167--172. IEEE.

\bibitem[{Cummins(2019)}]{cummins}
Cummins (2019).
\newblock How a turbocharger works.
\newblock
  \urlprefix\url{https://www.cummins.com/components/turbo-technologies/turbochargers/how-a-turbocharger-works}.

\bibitem[{Decombes et~al.(2002)Decombes, Pichouron, Maroteaux, Moreno, and
  Jullien}]{decombes2002simulation}
Decombes, G., Pichouron, J., Maroteaux, F., Moreno, N., and Jullien, J. (2002).
\newblock Simulation of the performance of a variable geometry turbocharger for
  diesel engine road propulsion.
\newblock \emph{International Journal of Thermodynamics}, 5(3), 139--149.

\bibitem[{Eriksson et~al.(2002)Eriksson, Nielsen, Brug{\aa}rd, Bergstr{\"o}m,
  Pettersson, and Andersson}]{eriksson2002modeling}
Eriksson, L., Nielsen, L., Brug{\aa}rd, J., Bergstr{\"o}m, J., Pettersson, F.,
  and Andersson, P. (2002).
\newblock Modeling of a turbocharged si engine.
\newblock \emph{Annual reviews in control}, 26(1), 129--137.

\bibitem[{Galindo et~al.(2014)Galindo, Tiseira, Fajardo, and
  Garc{\'\i}a-Cuevas}]{galindo2014development}
Galindo, J., Tiseira, A., Fajardo, P., and Garc{\'\i}a-Cuevas, L. (2014).
\newblock Development and validation of a radial variable geometry turbine
  model for transient pulsating flow applications.
\newblock \emph{Energy conversion and management}, 85, 190--203.

\bibitem[{Huang et~al.(2018)Huang, Ma, Li, Gao, and Qi}]{huang2018applying}
Huang, L., Ma, C., Li, Y., Gao, J., and Qi, M. (2018).
\newblock Applying neural networks (nn) to the improvement of gasoline
  turbocharger heat transfer modeling.
\newblock \emph{Applied Thermal Engineering}, 141, 1080--1091.

\bibitem[{Jadhav(2021)}]{suyash2021}
Jadhav, S. (2021).
\newblock \emph{Transient Modeling of Turbocharger Turbine using Physics and
  Machine Learning Techniques}.
\newblock Master's thesis, Purdue University.

\bibitem[{Jung and Glover(2005)}]{jung2005calibratable}
Jung, M. and Glover, K. (2005).
\newblock Calibratable linear parameter-varying control of a turbocharged
  diesel engine.
\newblock \emph{IEEE Transactions on control systems technology}, 14(1),
  45--62.

\bibitem[{K{\"a}chele(2020)}]{kachele2020turbocharger}
K{\"a}chele, A. (2020).
\newblock \emph{Turbocharger Integration into Multidimensional Engine
  Simulations to Enable Transient Load Cases}.
\newblock Springer.

\bibitem[{Kaiser et~al.(2020)Kaiser, Kutz, and Brunton}]{kaiser2020data}
Kaiser, E., Kutz, J.N., and Brunton, S.L. (2020).
\newblock Data-driven approximations of dynamical systems operators for
  control.
\newblock \emph{The Koopman Operator in Systems and Control}, 197--234.

\bibitem[{Korda and Mezi{\'c}(2018)}]{korda2018linear}
Korda, M. and Mezi{\'c}, I. (2018).
\newblock Linear predictors for nonlinear dynamical systems: Koopman operator
  meets model predictive control.
\newblock \emph{Automatica}, 93, 149--160.

\bibitem[{Kumar et~al.(2019)Kumar, Gope, Vijapur, and
  Nakayama}]{kumar2019hybrid}
Kumar, S., Gope, S., Vijapur, A., and Nakayama, S. (2019).
\newblock Hybrid plant modelling of diesel engine and after treatment systems
  using artificial neural networks.
\newblock Technical report, SAE Technical Paper.

\bibitem[{Kushwaha and Saraswati(2015)}]{kushwaha2015air}
Kushwaha, G. and Saraswati, S. (2015).
\newblock Air path identification of turbocharged diesel engine using rnn.
\newblock In \emph{2015 International Conference on Industrial Instrumentation
  and Control (ICIC)}, 1328--1332. IEEE.

\bibitem[{Lim(2016)}]{lim2016flow}
Lim, S.M. (2016).
\newblock \emph{Flow and heat transfer in a turbocharger radial turbine}.
\newblock Ph.D. thesis, KTH Royal Institute of Technology.

\bibitem[{Lusch et~al.(2018)Lusch, Kutz, and
  Brunton}]{lusch2018deep_machine_learning}
Lusch, B., Kutz, J.N., and Brunton, S.L. (2018).
\newblock Deep learning for universal linear embeddings of nonlinear dynamics.
\newblock \emph{Nature communications}, 9(1), 1--10.

\bibitem[{Ma et~al.(2019)Ma, Huang, and Vaidya}]{ma2019optimal}
Ma, X., Huang, B., and Vaidya, U. (2019).
\newblock Optimal quadratic regulation of nonlinear system using koopman
  operator.
\newblock In \emph{2019 American Control Conference (ACC)}, 4911--4916. IEEE.

\bibitem[{Mamakoukas et~al.(2021)Mamakoukas, Castano, Tan, and
  Murphey}]{mamakoukas2021derivative}
Mamakoukas, G., Castano, M.L., Tan, X., and Murphey, T.D. (2021).
\newblock Derivative-based koopman operators for real-time control of robotic
  systems.
\newblock \emph{IEEE Transactions on Robotics}, 37(6), 2173--2192.

\bibitem[{Mandic and Chambers(2001)}]{mandic2001recurrent}
Mandic, D. and Chambers, J. (2001).
\newblock \emph{Recurrent neural networks for prediction: learning algorithms,
  architectures and stability}.
\newblock Wiley.

\bibitem[{Marelli et~al.(2017)Marelli, Gandolfi, and
  Capobianco}]{marelli2017heat}
Marelli, S., Gandolfi, S., and Capobianco, M. (2017).
\newblock Heat transfer effect on performance map of a turbocharger turbine for
  automotive application.
\newblock Technical report, SAE Technical Paper.

\bibitem[{Moraal and Kolmanovsky(1999)}]{moraal1999turbocharger}
Moraal, P. and Kolmanovsky, I. (1999).
\newblock Turbocharger modeling for automotive control applications.
\newblock Technical report, SAE Technical Paper.

\bibitem[{Proctor et~al.(2018)Proctor, Brunton, and
  Kutz}]{proctor2018generalizing_koopman_control_3}
Proctor, J., Brunton, S., and Kutz, J. (2018).
\newblock Generalizing \textsc{K}oopman theory to allow for inputs and control.
\newblock \emph{SIAM Journal on Applied Dynamical Systems}, 17(1), 909--930.

\bibitem[{Proctor et~al.(2016)Proctor, Brunton, and Kutz}]{proctor2016dynamic}
Proctor, J.L., Brunton, S.L., and Kutz, J.N. (2016).
\newblock Dynamic mode decomposition with control.
\newblock \emph{SIAM Journal on Applied Dynamical Systems}, 15(1), 142--161.

\bibitem[{Pulpeiro~Gonzalez et~al.(2019)Pulpeiro~Gonzalez, Ankobea-Ansah,
  Escuder~Milian, and Hall}]{pulpeiro2019modeling}
Pulpeiro~Gonzalez, J., Ankobea-Ansah, K., Escuder~Milian, E., and Hall, C.M.
  (2019).
\newblock Modeling the gas exchange processes of a modern diesel engine with an
  integrated physics-based and data-driven approach.
\newblock In \emph{Dynamic Systems and Control Conference}, volume 59155,
  V002T11A004. American Society of Mechanical Engineers.

\bibitem[{Serrano et~al.(2008)Serrano, Arnau, Dolz, Tiseira, and
  Cervell{\'o}}]{serrano2008model}
Serrano, J., Arnau, F., Dolz, V., Tiseira, A., and Cervell{\'o}, C. (2008).
\newblock A model of turbocharger radial turbines appropriate to be used in
  zero-and one-dimensional gas dynamics codes for internal combustion engines
  modelling.
\newblock \emph{Energy Conversion and Management}, 49(12), 3729--3745.

\bibitem[{Wang(2008)}]{wang2008hybrid}
Wang, J. (2008).
\newblock Hybrid robust air-path control for diesel engines operating
  conventional and low temperature combustion modes.
\newblock \emph{IEEE Transactions on Control Systems Technology}, 16(6),
  1138--1151.

\bibitem[{Williams et~al.(2016)Williams, Hemati, Dawson, Kevrekidis, and
  Rowley}]{williams2016extending_koopman_control_4}
Williams, M.O., Hemati, M.S., Dawson, S.T., Kevrekidis, I.G., and Rowley, C.W.
  (2016).
\newblock Extending data-driven \textsc{K}oopman analysis to actuated systems.
\newblock \emph{IFAC-PapersOnLine}, 49(18), 704--709.

\bibitem[{Williams et~al.(2015)Williams, Kevrekidis, and
  Rowley}]{williams2015data_koopman_accuracy_1}
Williams, M.O., Kevrekidis, I.G., and Rowley, C.W. (2015).
\newblock A data--driven approximation of the \textsc{K}oopman operator:
  Extending dynamic mode decomposition.
\newblock \emph{Journal of Nonlinear Science}, 25(6), 1307--1346.

\bibitem[{Yeung et~al.(2019)Yeung, Kundu, and Hodas}]{yeung2019learning}
Yeung, E., Kundu, S., and Hodas, N. (2019).
\newblock Learning deep neural network representations for koopman operators of
  nonlinear dynamical systems.
\newblock In \emph{2019 American Control Conference (ACC)}, 4832--4839. IEEE.

\end{thebibliography}

\end{document}